\documentclass{ieeeaccess}
\usepackage{inputenc}
\usepackage{fontenc}
\usepackage{mathtext}
\usepackage{textcase}
\usepackage{bm}
\usepackage{amssymb,amsfonts}
\usepackage{subfigure}
\usepackage{mathrsfs}
\usepackage{color}
\usepackage{textcomp}
\usepackage{multirow}
\usepackage{cite}
\usepackage{graphicx}
\usepackage{epstopdf}
\usepackage{caption}
\usepackage{algorithm}

\usepackage{algpseudocode}
\usepackage{amsmath}
\def\BibTeX{{\rm B\kern-.05em{\sc i\kern-.025em b}\kern-.08em
    T\kern-.1667em\lower.7ex\hbox{E}\kern-.125emX}}
\begin{document}
\history{Date of publication xxxx 00, 0000, date of current version xxxx 00, 0000.}
\doi{10.1109/ACCESS.2017.DOI}

\title{Multi-Antenna Joint Radar and Communications: Precoder Optimization and Weighted Sum-Rate vs Probing Power Tradeoff}
\author{\uppercase{Chengcheng Xu}\authorrefmark{1}, 
\uppercase{Bruno Clerckx\authorrefmark{2},\IEEEmembership{Senior Member, IEEE}, and Jianyun Zhang\authorrefmark{1},
\IEEEmembership{Member, IEEE}}}
\address[1]{College of Electronic Engineering, National University of Defense Technology, Hefei 230037, China}
\address[2]{Department of Electrical and Electronic Engineering, Imperial College London, London SW7 2AZ,  United Kingdom (e-mail: b.clerckx@imperial.ac.uk)}
\tfootnote{}

\markboth
{Author \headeretal: Preparation of Papers for IEEE TRANSACTIONS and JOURNALS}
{Author \headeretal: Preparation of Papers for IEEE TRANSACTIONS and JOURNALS}

\corresp{Corresponding author: Chengcheng Xu (e-mail: xuchengcheng17@nudt.edu.cn).}

\begin{abstract}
In order to further exploit the potential of joint multi-antenna radar-communication (RadCom) system, we propose two transmission techniques respectively based on separated and shared antenna deployments. Both techniques are designed to maximize weighted sum rate (WSR) and probing power at target's location under average power constraints at antennas such that the system can simultaneously communicate with downlink users and detect the target within the same frequency band. Based on a Weighted Minimized Mean Square Errors (WMMSE) method, the separated deployment transmission is designed via semidefinite programming (SDP) while the shared deployment problem is solved by majorization-minimization (MM) algorithm. Numerical results show that shared deployment outperforms separated deployment in radar beamforming. The tradeoffs between WSR and probing power at target are compared among both proposed transmissions and two practically simpler dual-function implementations i.e., time division and frequency division. Results show that although separated deployment has an advantage of realizing spectrum sharing, it experiences a performance loss compared with frequency division, while shared deployment outperforms both and surpasses time division in certain conditions.
\end{abstract}


\titlepgskip=-15pt

\maketitle

\section{Introduction}
The 4th and 5th generation wireless communication systems are competing with long-range radar applications in the S-band (2-4GHz) and C-band (4-8GHz), which will possibly result in severe spectrum congestion and hamper the higher data rate requirements for the increasing demand in future wireless communication\cite{6967722}. Though efforts for new spectrum management regulations and policies are needed, a longer-term solution is to enable communication and radar spectrum sharing (CRSS). There are two main research topics in the field of CRSS: 1) coexistence of existing radar and communication devices, 2) co-design for dual-function systems. \par 
\subsection{Coexistence of Existing Radar and Communication Devices} 
For coexistence of existing radar and communication devices, research focuses on designing high-quality wideband radar waveforms that achieve spectrum nulls on communication frequency bands\cite{Aubry2016}. On this basis, \cite{rowe2014spectrally,8528529} then design waveforms with more accurate spectrum shapes for higher spectrum efficiency. However, instead of designing the radar waveforms only, \cite{Zheng2018}  jointly designs communication precoders together with the slow-time radar coding waveforms to ensure both radar Signal-to-Interference-plus-Noise (SINR) and communication rate requirements. Nevertheless, all the aforementioned works are limited to single-antenna radar systems. As multi-antenna processing can greatly improve radar performance\cite{li2007mimo}, \cite{Tang2019MIMOspec} extends the spectral constraint towards Multiple-Input Multiple-Output (MIMO) radar waveform design and enables MIMO radar to work in a spectrally crowded environment. In contrast, \cite{liu2018power} designs the precoder of the multi-user MIMO (MU-MIMO) communication base station (BS) to coexist with the MIMO radar. Instead of designing the radar or communication system solely, \cite{li2017joint} and \cite{qian2018joint} have been devoted to the coordinated design of both existing MIMO communication systems and MIMO radar systems to achieve coexistence. 
Given the existing infrastructure, a coexistence approach manages interference between radar and communication as much as it can. However, for uncoordinated coexistence design, some important phenomena are not considered in the simplified scenarios \cite{Zheng2019Overview}, while for coordinated coexistence, governmental and military agencies might be unwilling to upgrade the existing deployment \cite{Liu2019HOW}. \par 
\subsection{Dual-function System Design} 
Accounting for the possible drawbacks aforementioned, designing a dual-function system that makes the best use of the spectrum for both detecting and communicating might be a better alternative. Early studies \cite{saddik2007ultra,sturm2011waveform} consider single-antenna dual-function platforms without utilizing multi-antenna processing. Then, based on the waveform diversity of MIMO radar and the concept of space-division multiple access (SDMA) in MIMO communication, \cite{hassanien2015dual,hassanien2016phase} embed the information stream into radar pulses via a multi-antenna platform, detecting targets at the mainlobe and transmitting information streams at sidelobe. The sidelobe level is modulated via amplitude shift keying (ASK), where different powers correspond to different communication symbols in \cite{hassanien2015dual}. Likewise, \cite{hassanien2016phase} also develops phase shift keying (PSK) in this system by representing the symbols as the different phases of the signals received at the angle of the sidelobe. One significant restriction of such a dual-function system is that the rate is limited by the Pulse Repetition Frequency (PRF), which is far from satisfactory for communication requirements. To overcome this problem, \cite{liu2018mu} proposes a joint multi-antenna radar-communication (RadCom) system defined as a dual-function platform simultaneously transmitting probing signals to radar targets and serving multiple downlink users. Both functions are realized within the same frequency band. Specifically, two antenna deployments are mentioned. Separated deployment splits the antennas into two groups respectively working as MIMO radar and BS, while the shared deployment only transmits communication streams and the precoders are designed to form a desired radar beampattern and meet the SINR requirements for communication users. However, only communication SINR is adopted as a metric, but a more representative communication performance metric such as rate is not considered in this work. In addition, the performance of joint RadCom has not been compared with practically simpler implementation using orthogonal resources in time or frequency to fulfill the dual function, which is an essential criterion to decide where joint RadCom is worth the efforts. \par 
\subsection{Contribution}
In this paper, we propose two multi-antenna RadCom transmission design techniques based on separated and shared antenna deployments respectively. Both techniques enable the platform to simultaneously communicate with downlink users and probe one target of interest within the same frequency band. Major contributions are summarized as follows.
\begin{enumerate}
\item{We propose transmission techniques that maximize the weighted sum rate (WSR) of communication and the probing power at target's location for both separated and shared deployments.}\par  Since WSR is the most representative metric of a communications system, we consider WSR maximization instead of SINR constraint at each user in \cite{liu2018mu}. To the best of our knowledge, we are the first to consider WSR maximization in the system model with precoders. We also consider probing power maximization at the target's location rather than turning this metric into beampattern approximation problem in \cite{liu2018mu}. This makes our transmission design more direct for typical MIMO radar tracking and scanning mode, and enables a more clear tradeoff comparison. However, by adopting WSR and probing power at target, the transmission design problem becomes difficult. Specifically, maximizing WSR as a sum of logarithms and probing power as a quadratic form under power constraints makes the optimization problem highly non-convex and intractable.
\item{We propose WMMSE-SDP and WMMSE-MM algorithms respectively to solve the two proposed transmission design problems.}\par 
 For separated deployment, we propose to reformulate the problem into semidefinite programming (SDP) based on Weighted Minimized Mean Square Errors (WMMSE) method. For the shared deployment, non-convex per-antenna power constraint makes the design problem even more difficult. We propose a majorization-minimization (MM) iterative algorithm based on WMMSE to effectively solve the problem.
\item{We compare the performance of the proposed transmission techniques with practically simpler time-division and frequency-division dual-function implementations.} \par 
In order to provide a well-rounded evaluation of our proposed techniques, we compare the tradeoffs of both separated and shared multi-antenna RadCom deployments with time-division and frequency-division dual-function implementation which might also be practical options because of plain and easy realization. Separated deployment has an advantage of realizing spectrum sharing compared with frequency division, but is surpassed by the latter in tradeoff performance. In contrast, the shared deployment outperforms frequency division with a significant tradeoff gain, and exceeds time division in certain conditions. 
	
\end{enumerate}
\subsection{Organization}
The rest of the paper is organized as follows. The system models and metrics of both separated and shared deployments are illustrated in Section II. In Section III, optimization problems of transmission designs for both deployments are formulated. Algorithms for solving the optimization problems are subsequently presented in Section IV. Section V demonstrates the simulation results and analysis. Section VI concludes the paper.
\section{System Model and metrics}
In this work, we adopt the separated and shared deployment models of \cite{liu2018mu}, where either deployment works simultaneously as a BS serving downlink users and a collocated MIMO radar probing the target of interest. Both deployments are equipped with a total of $N_{\text{t}}$ antennas, serving $K$ single-antenna users indexed as $\mathcal{K} =\{1,\dots, K\}$. Typically, we assume that both deployments use a uniform linear array (ULA) in our system model. The total power budget for either deployment is $P_{\text{t}}$. We assume that the RadCom system works in a tracking mode as a radar, where there is typically one target of interest at the azimuth angle of $\theta_m$ \cite{onTransMIMO}. Beamforming is thus expected.\par
\begin{figure}[htpb]
	\centering
	\subfigure[Schematic diagram for separated deployment] {\includegraphics[width=0.45\linewidth]{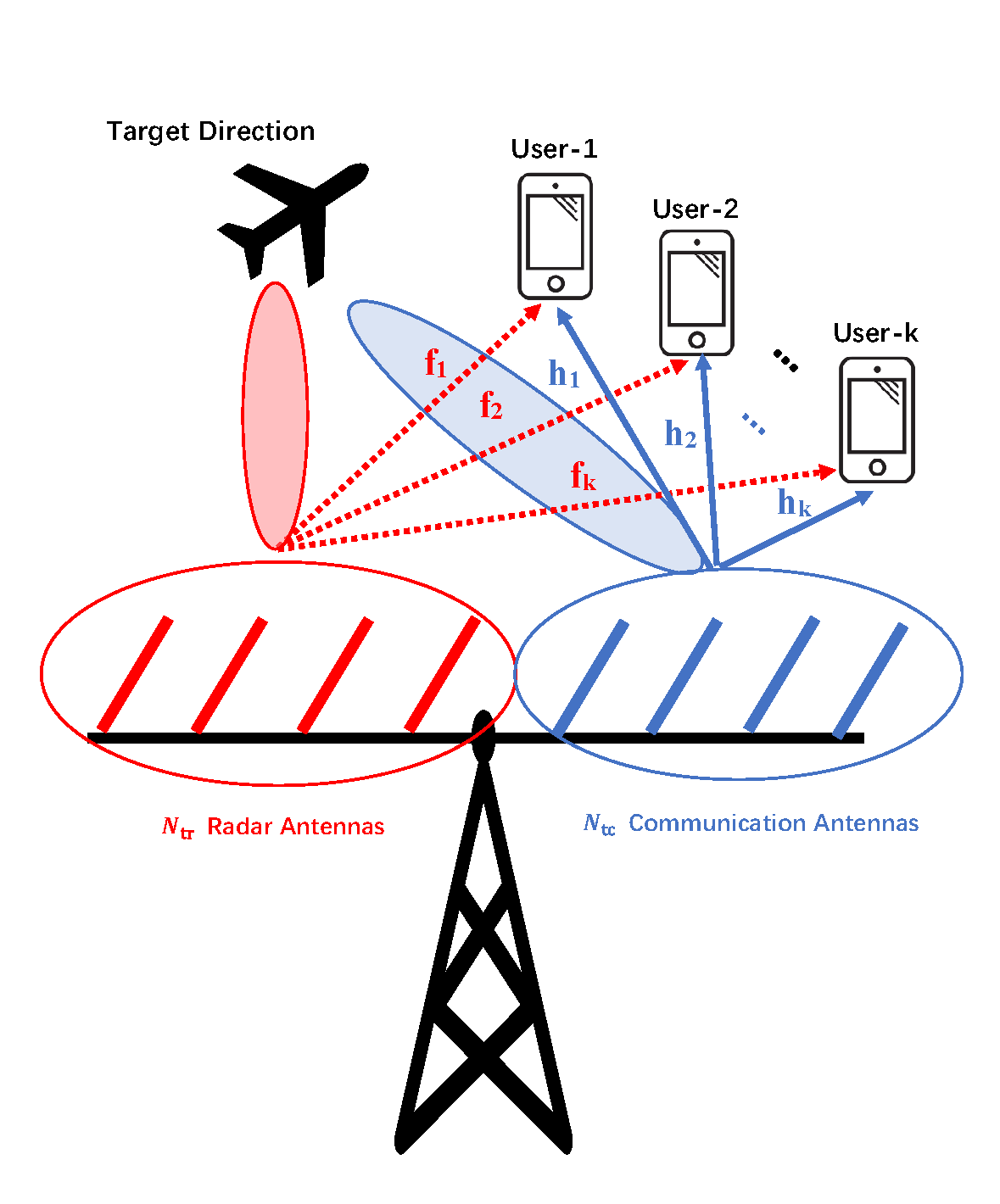}\label{SEP}}
	\subfigure[Schematic diagram for shared deployment] {\includegraphics[width=0.45\linewidth]{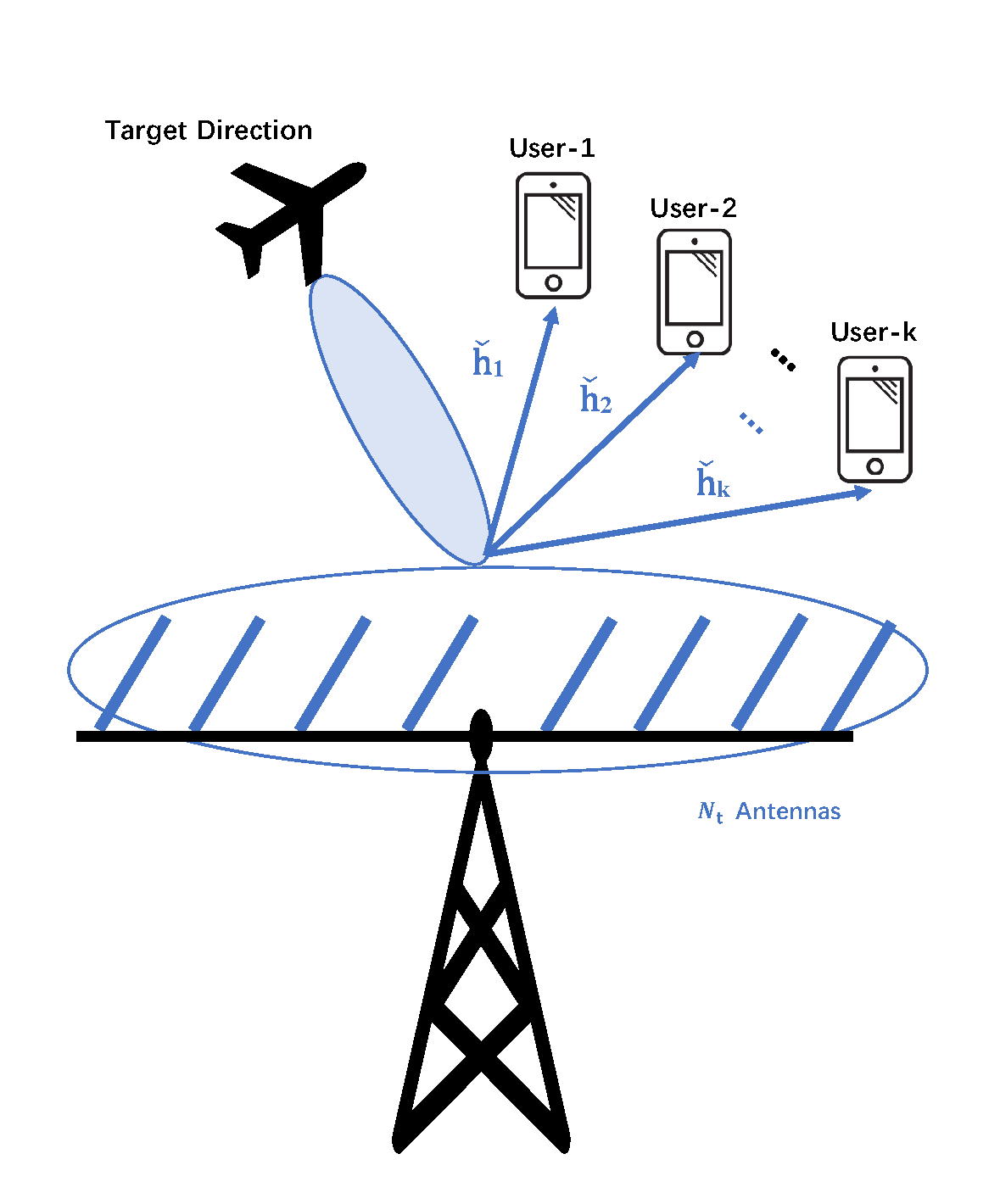}\label{SHA}}
	\caption{Schematic diagram for separated and shared multi-antenna joint RadCom}\label{SD}
\end{figure} 
\subsection{Separated deployment}
The separated deployment splits the antennas into two groups, i.e., a group of $N_{\text{tr}}$ antennas only transmitting radar signals and the other group of $N_{\text{tc}}$ antennas only transmitting communication signals. Both the communication precoders and the radar signals are designed to fulfill the dual function. The schematic diagram is shown in Fig. \ref{SEP}.\par 
The received signal at user-$k$ can be expressed as 
\begin{equation}
y^{\text{S}}_k[l]={\bf h}_k^H\sum_{j\in \mathcal{K}}{\bf p}_j{s}_j[l]+{\bf f}_{k}^H{\bf r}_{l}+n_k[l]
\end{equation} 
where ${\bf h}_k\in\mathbb{C}^{N_{\text{tc}}\times 1}$ and ${\bf f}_k\in\mathbb{C}^{N_{\text{tr}}\times 1}$ are respectively the non-line-of-sight (NLOS) channel vectors from communication antennas and radar antennas to user-$k$. $s_j[l]$ and $n_j[l]\sim \mathcal{CN}(0, \sigma_n^2)$ are the communication symbols and receiving noise of user-$j$ at the time index $l$. Without loss of generality, we assume $\sigma_n^2=1$. ${\bf p}_j\in\mathbb{C}^{N_{\text{tc}}\times 1}$ is the precoder for user-$j$, and ${\bf r}_{l}\in\mathbb{C}^{N_{\text{tr}}\times 1}$ is the $l$th snapshot of radar antennas. The covariance matrix of transmit radar signal is ${\bf R}_{\text{x}}=\frac{1}{L}{\bf r}_{l}{\bf r}_{l}^H$ with $L$ be the length of signal on fast-time axis. \par 
We first introduce WSR as the communication metric. The SINR of decoding $s_k$ at user-$k$ is 
\begin{equation}
	\gamma^{\text{S}}_{k}({\bf P},{\bf R}_{\text{x}})=\frac{\left| {\bf h}_k^H{\bf p}_{k}\right| ^2}{\sum_{j\in \mathcal{K},j\neq k}\left| {\bf h}_k^H{\bf p}_j\right|^2+{\bf f}_k^H{\bf R}_{\text{x}}{\bf f}_k+1},\forall k \in \mathcal{K}\label{sinr}
\end{equation}
where ${\bf P}=[{\bf p}_1,\dots,{\bf{p}}_K]$ is the precoder matrix of the separated deployment. Therefore, the achievable rate at user-$k$ in the separated deployment can be denoted as
\begin{equation}
	R^{\text{S}}_{k}({\bf P},{\bf R}_{\text{x}})=\log_2\left(1+\gamma^{\text{S}}_{k}({\bf P},{\bf R}_{\text{x}})\right).
\end{equation} 
Denoting the rate weight of user-$k$ as $\mu_k$, WSR of the separated deployment is $\sum_{k\in \mathcal{K}}\mu_kR^{\text{S}}_{k}({\bf P},{\bf R}_{\text{x}})$.\par
Then, we consider our dual-function system works as a MIMO radar where the target location is known or estimated and we aim at maximizing cumulated power of the probing signals at the target location $\theta_m$. Note that this is a simple and classic MIMO radar beamforming scenario illustrated in \cite{stoica2007probing}. The probing power is  
\begin{equation}
	P^{\text{S}}_{\text{T}}(\theta_m)={\bf a}^H(\theta_m){\bf C}_{\text{t}}{\bf a}(\theta_m)\label{ptor}
\end{equation}
where ${\bf a}(\theta_m)=[1,e^{j2\pi\delta sin(\theta_m)},\dots,e^{j2\pi(N_{\text{t}}-1)\delta sin(\theta_m)}]^T\in\mathbb{C}^{N_{\text{t}}\times 1}$ is the transmit steering vector of ULA, and $\delta$ is the normalized distance (relative to wavelength) between adjacent array elements. For other array structures, the expression of ${\bf a}(\theta_m)$ needs to be changed. ${\bf C}_{\text{t}}\in\mathbb{C}^{N_{\text{t}}\times N_{\text{t}}}$ is the overall transmit covariance matrix. Assuming the radar signals are statistically independent with communication signals, we have
\begin{equation}
	{\bf C}_{\text{t}}=\left[\begin{array}{cc}
	{\bf R}_{\text{x}}  &{\bf 0}\\
	 {\bf 0} &{\bf P}{\bf P}^H
	\end{array}\right].
\end{equation} 
Thus \eqref{ptor} is reformulated as $P^{\text{S}}_{\text{T}}(\theta_m)={\bf a}_1^H(\theta_m){\bf R}_{\text{x}}{\bf a}_1(\theta_m)+{\bf a}_2^H(\theta_m){\bf P}{\bf P}^H{\bf a}_2(\theta_m)$ where ${\bf a}_1(\theta_m)\in\mathbb{C}^{N_{\text{tr}}\times1}$ and ${\bf a}_2(\theta_m)\in\mathbb{C}^{N_{\text{tc}}\times1}$ satisfy ${\bf a}(\theta_m)=[{\bf a}_1(\theta_m);{\bf a}_2(\theta_m)]$.\par 
 
\subsection{Shared Deployment}
For the shared deployment, $N_{\text{t}}$ antennas all transmit precoded communication streams only and fulfill the dual function. The schematic diagram is shown in Fig. \ref{SHA}. In this deployment, only the precoders are designed. The received signal at user-$k$ is 
\begin{equation}
	y^{\text{J}}_k[l]=\check{\bf h}_k^H\sum_{j\in \mathcal{K}}\check{\bf p}_j{s}_j[l]+n_k[l]
\end{equation}
where $\check{\bf h}_k\in\mathbb{C}^{N_{\text{t}}\times 1}$ is the channel vector between the shared system and user-$k$. Differently, without the interference caused by radar signals, the SINR at user-$k$ is 
\begin{equation}
\gamma^{\text{J}}_{k}(\check{\bf P})=\frac{\left| \check{\bf h}_k^H\check{\bf p}_{k}\right| ^2}{\sum_{j\in \mathcal{K},j\neq k}\left| \check{\bf h}_k^H\check{\bf p}_j\right|^2+1},\forall k \in \mathcal{K}
\end{equation}
where $\check{\bf P}=[\check{\bf p}_1,\dots,\check{\bf{p}}_K]$ is the precoder matrix for the shared deployment. Thus, the rate at user-$k$ is
\begin{equation}
R^{\text{J}}_{k}(\check{\bf P})=\log_2\left(1+\gamma^{\text{J}}_{k}(\check{\bf P})\right)
\end{equation}  
and WSR of the shared deployment is $\sum_{k\in \mathcal{K}}\mu_kR^{\text{S}}_{k}({\bf P},{\bf R}_{\text{x}})$.\par 
In the shared deployment, probing power at the target location $\theta_m$ is $P^{\text{J}}_{\text{T}}(\theta_m)={\bf a}^H(\theta_m)\check{\bf P}\check{\bf P}^H{\bf a}(\theta_m)$. Although we do not focus on the radar matched filter design in this work, the transmit signal of shared deployment coincides with the transmit model of one collocated MIMO radar in \cite{onTransMIMO} where the matched filter settings for radar detection can be referred to as an option. It is also derived in \cite{onTransMIMO} that maximizing output signal-to-noise-ratio (SNR) of the detector is equivalent to maximizing $P^{\text{J}}_{\text{T}}(\theta_m)$.
\section{Problem Formulation}
The transmission design problem for the separated deployment can be expressed as 
\begin{subequations}\label{seperated problem}
\begin{align}
\max\limits_{{\bf P},{\bf R}_{\text{x}}} \quad &\rho\sum_{k\in \mathcal{K}}\mu_kR^{\text{S}}_{k}({\bf P},{\bf R}_{\text{x}})+{\bf a}_1^H(\theta_m){\bf R}_{\text{x}}{\bf a}_1(\theta_m)\notag\\&+{\bf a}_2^H(\theta_m){\bf P}{\bf P}^H{\bf a}_2(\theta_m)\label{objsep}\\ 
s.t.\quad&\text{diag}({\bf R}_{\text{x}})=\frac{P_{\text{r}}{\bf 1}^{N_{\text{tr}}\times 1}}{N_{\text{tr}}}\label{radpower}\\
&\text{tr}({\bf P}{\bf P}^H)\le P_{\text{c}}\label{compower}\\
&{\bf R}_{\text{x}}\succeq 0\label{sd}
\end{align}
\end{subequations}
where ${\bf P}$ is the precoders of communication streams, ${\bf R}_{\text{x}}$ is the transmit covariance matrix of radar signals, $P_{\text{c}}$ and $P_{\text{r}}$ are the transmit power budgets of radar and communication sub-arrays respectively. The first counterpart of the objective function \eqref{objsep} represents WSR while the rest parts denote probing power at the target. Both metrics are maximized via regularization with a parameter $\rho$. Although communication and radar signals are separately transmitted by two sub-systems, they show mutual effects on each other when operating simultaneously, i.e., radar signals cause interference to communication users and communication signals are supposed to help probe the target. Constraint \eqref{radpower} is the uniform elemental power constraint in radar implementation \cite{stoica2007probing}, and \eqref{compower} is the total power constraint in communication implementation. \eqref{sd} restricts ${\bf R}_{\text{x}}$ to be semi-definite.\par 
However, the dual function can also be fulfilled by the shared deployment, of which the transmission design problem can be formulated as
\begin{equation}\label{joint origin}
\begin{split}
\max\limits_{\check{\bf P}} \quad &\rho\sum_{k\in \mathcal{K}}\mu_kR^{\text{J}}_{k}(\check{\bf P})+{\bf a}^H(\theta_m)\check{\bf P}\check{\bf P}^H{\bf a}(\theta_m)\\ 
s.t.\quad&\text{diag}(\check{\bf P}\check{\bf P}^H)=\frac{P_{\text{t}}{\bf 1}^{N_{\text{t}}\times 1}}{N_{\text{t}}}.
\end{split}
\end{equation}
Likewise, WSR and probing power maximization are combined in the objective function via regularization. There is also an elemental power constraint for all antennas. The total power budget constraint for communication is omitted because it is certainly satisfied when the elemental power restriction is met.  \par 
It needs pointing out that the maximum probing power design approach used in \eqref{sd} and \eqref{joint origin} might have drawbacks when extended to the scenario of multiple targets according to \cite{stoica2007probing}.
\section{WMMSE-based Solving Algorithms}
It is clear that both separated and shared transmission design problems \eqref{seperated problem} and \eqref{joint origin} are non-convex because of the intractable form of WSR and maximizing a quadratic power function in objective functions. However, this problem can be reformulated using the WMMSE approach and solved through the WMMSE-based Alternating Optimization (WMMSE-AO) algorithm following \cite{mao2019RSWIPT}.\par 
\subsection{WMMSE-SDP algorithm for separated transmission}
We decode $s_{k}$ at user-$k$ via an equalizer $g_{k}$, and get the estimation $\hat{s}_k$ of $s_k$ as $\hat{s}_{k}=g_{k} y^{\text{S}}_k$.
Subsequently, the Mean Square Errors (MSE) of estimation, defined as $\varepsilon_{k}\triangleq\mathbb{E}\left\{\left| \hat{s}_k-s_k\right|^2\right\}$, can be expressed as $\varepsilon_{k}=\left|g_{k}\right|^2T_{k}-2\mathfrak{R}\left\{g_{k}{\bf h}_k^H{\bf p}_k\right\}+1$
where 
\begin{equation}
T_{k}\triangleq \sum_{j\in \mathcal{K}}\left|{\bf h}_k^H{\bf p}_j\right|^2+\text{tr}({\bf R}_{\text{x}}{\bf f}_k{\bf f}_k^H)+1.
\end{equation}
Optimum equalizers are obtained by letting  $\frac{\partial\varepsilon_{k}}{\partial g_{k}}=0$, which are also the MMSE equalizers given by 
\begin{equation}
g^{\text{MMSE}}_{k}={\bf p}_k^H{\bf h}_k(T_{k})^{-1}.
\end{equation}
Minimized MSEs (MMSEs) based on $g^{\text{MMSE}}_{k}$ are given by
\begin{equation}
\varepsilon^{\text{MMSE}}_{k}\triangleq \min\limits_{g_{k}}\varepsilon_{k}=(T_{k})^{-1}\left(T_{k}-\left|{\bf h}_k^H{\bf p}_k\right|^2\right).
\label{MMSEs}
\end{equation}\par 
Hence, by comparing \eqref{MMSEs} with \eqref{sinr}, we rewrite SINRs of decoding the intended streams at user-$k$ as  $\gamma^{\text{S}}_{k}=(1/\varepsilon_{k}^{\text{MMSE}})-1$, and the rate as  $R^{\text{S}}_{k}=\log_2(1+\gamma_{k})=-\log_2(\varepsilon_{k}^{\text{MMSE}})$.\par 
By allocating a positive weight $w_k$ to user-$k$'s rate, we define the augmented WMSEs as $\xi_{k}\triangleq w_{k}\varepsilon_{k}-\log_2(w_{k})$.
After optimizing over the equalizers and weights, the Rate-WMMSE relationships are
\begin{equation}
\xi_{k}^{\text{MMSE}}\triangleq\min\limits_{w_{k},g_{k}}\xi_{k}=1-R^{\text{S}}_{k}
\end{equation}
where the optimum equalizers and the optimum weights are  
\begin{equation}
\begin{split}\label{updatewg}
&g_k^{*}=g_k^{\text{MMSE}}\\
&w_{k}^{*}=w_{k}^{\text{MMSE}}=\left(\varepsilon_{k}^{\text{MMSE}}\right)^{-1},
\end{split}
\end{equation}
resulting from meeting the first order optimality conditions.\par 
Using the rate-WMMSE relationships, we can then reformulate \eqref{seperated problem} as 
\begin{subequations}\label{SepWMMSE}
	\begin{align}
	\min\limits_{{\bf P},{\bf R}_{\text{x}},{\bf w},{\bf g}} \quad &\rho\sum_{k\in \mathcal{K}}\mu_k\xi_{k}({\bf P},{\bf R}_{\text{x}})-{\bf a}_2^H(\theta_m)\left({\bf P}{\bf P}^H\right){\bf a}_2(\theta_m)\notag\\&-{\bf a}_1^H(\theta_m){\bf R}_{\text{x}}{\bf a}_1(\theta_m)\label{objSepWMMSE}\\ 
	s.t.\quad&\text{diag}({\bf R}_{\text{x}})=\frac{P_{\text{r}}{\bf 1}^{N_{\text{tr}}\times 1}}{N_{\text{tr}}/2}\\
	&\text{tr}({\bf P}{\bf P}^H)\le P_{\text{c}}\\
	&{\bf R}_{\text{x}}\succeq 0
	\end{align}
\end{subequations}
where ${\bf w}=\left[w_1,w_1,\dots,w_K\right]$ is the vector of all MSE weights. ${\bf g}=\left[g_1,g_2,\dots,g_K\right]$ is the vector of all equalizers. It is worth noting that the second term $-{\bf a}_2^H(\theta_m)\left({\bf P}{\bf P}^H\right){\bf a}_2(\theta_m)$ in \eqref{objSepWMMSE} is non-convex. To make it convex, we first reformulate this part as\par
\begin{equation}
\begin{split}
&-{\bf a}_2^H(\theta_m)\left({\bf P}{\bf P}^H\right){\bf a}_2(\theta_m)=-\sum_{k\in \mathcal{K}}{\bf p}_k^H{\bf a}_2(\theta_m){\bf a}_2^H(\theta_m){\bf p}_k\\
&=N_{\text{tc}}\times P_{\text{c}}-\sum_{k\in \mathcal{K}}{\bf p}_k^H{\bf a}_2(\theta_m){\bf a}_2^H(\theta_m){\bf p}_k-N_{\text{tc}}\times P_{\text{c}}\\
&=\sum_{k\in \mathcal{K}}{\bf p}_k^H(N_{\text{tc}}{\bf I}){\bf p}_k-\sum_{k\in \mathcal{K}}{\bf p}_k^H{\bf a}_2(\theta_m){\bf a}_2^H(\theta_m){\bf p}_k-N_{\text{tc}}\times P_{\text{c}}\\
&=\sum_{k\in \mathcal{K}}{\bf p}_k^H\left(N_{\text{tc}}{\bf I}-{\bf a}_2(\theta_m){\bf a}_2^H(\theta_m)\right){\bf p}_k-N_{\text{tc}}\times P_{\text{c}}\\
&=\sum_{k\in \mathcal{K}}{\bf p}_k^H{\bf Z}(\theta_m){\bf p}_k-N_{\text{tc}}\times P_{\text{c}}
\end{split} 
\end{equation}

where we denote ${\bf Z}(\theta_m)=N_{\text{tc}}{\bf I}-{\bf a}_2(\theta_m){\bf a}_2^H(\theta_m)$. For the definition of steering vector in \eqref{ptor}, it is clear that ${\bf a}_2(\theta_m){\bf a}_2^H(\theta_m)$ is a rank-1 matrix with the eigenvalue of $\lVert{\bf a}_2(\theta_m)\lVert^2=N_{\text{tc}}$. Therefore, ${\bf Z}(\theta_m)$ is semi-definite. By omitting the constant part, we have that minimizing $-{\bf a}_2^H(\theta_m)\left({\bf P}{\bf P}^H\right){\bf a}_2(\theta_m)$ is equivalent to minimizing $\sum_{k\in \mathcal{K}}{\bf p}_k^H{\bf Z}(\theta_m){\bf p}_k$, which is convex.\par 
Afterwards, \eqref{SepWMMSE} is equivalent to 
\begin{equation}\label{SepWMMSEfinal}
	\begin{split}
	\min\limits_{{\bf P},{\bf R}_{\text{x}},{\bf w},{\bf g}} \quad &\rho\sum_{k\in \mathcal{K}}\mu_k\xi_{k}({\bf P},{\bf R}_{\text{x}})+\text{tr}\left({\bf Z}(\theta_m){\bf P}{\bf P}^H\right)\\&-\text{tr}\left({\bf R}_{\text{x}}{\bf a}_1(\theta_m){\bf a}_1^H(\theta_m)\right)\\ 
	s.t.\quad&\text{diag}({\bf R}_{\text{x}})=\frac{P_{\text{r}}{\bf 1}^{N_{\text{tr}}\times 1}}{N_{\text{tr}}}\\
	&\text{tr}({\bf P}{\bf P}^H)\le P_{\text{c}}\\
	&{\bf R}_{\text{x}}\succeq 0.
	\end{split}
\end{equation}
Note that when $\{{\bf w},{\bf g}\}$ are fixed, \eqref{SepWMMSEfinal} is a semidefinite programming (SDP) convex problem that can be efficiently solved by CVX toolbox, and optimum $\{{\bf w}^*,{\bf g}^*\}$ can be updated following \eqref{updatewg}. Therefore, we here use the WMMSE-based AO algorithm with details in \cite{mao2019RSWIPT} to solve the problem, which is summarized in Algorithm 1. After having the optimum ${\bf R}_{\text{x}}$, the radar snapshots can be further obtained using algorithms in \cite{Luo2010Semidefinite}.

\begin{algorithm}[h]
	\caption{WMMSE-SDP algorithm}
	\begin{algorithmic}[1]
	\Require  $t\leftarrow0$, ${\bf P}^{[t]}$
    \State$\text{WSR}^{[t]}$ is calculated from ${\bf P}^{[t]}$ 
	\Repeat
	\State	${\bf w}^{*}\leftarrow{\bf w}^{\text{MMSE}}({\bf P}^{[t]})$;
		\State	${\bf g}^{*}\leftarrow{\bf g}^{\text{MMSE}}({\bf P}^{[t]})$;
		\State	update ${\bf P}^{[t+1]}$, ${\bf R}^{[t+1]}_{\text{x}}$ by solving SDP \eqref{SepWMMSEfinal} with updated ${\bf w}^{*},{\bf g}^{*}$;
		\State	update $\text{WSR}^{[t+1]}$ using ${\bf P}^{[t+1]}$.
		\State	$t++$;
	\Until{$\vert\text{WSR}^{[t]}-\text{WSR}^{[t-1]}\vert\leq\epsilon_1$}
	\end{algorithmic}
\end{algorithm}
\subsection{WMMSE-MM algorithm for shared transmission}
For the shared transmission design problem, we first follow the same path in Section IV.A and reformulate \eqref{joint origin} with WMMSE method. To simplify, we omit the repetitive parts and directly give the reformulated problem as
\begin{equation}
\begin{split}\label{WMMSE}
\min\limits_{\check{\bf P},\check{{\bf w}},\check{{\bf g}}} \quad &\rho\sum_{k\in \mathcal{K}}\mu_k\zeta_{k}(\check{\bf P})-{\bf a}^H(\theta_m)\check{\bf P}\check{\bf P}^H{\bf a}(\theta_m)\\ 
s.t.\quad&\text{diag}(\check{\bf P}\check{\bf P}^H)=\frac{P_{\text{t}}{\bf 1}^{N_{\text{t}}\times 1}}{N_{\text{t}}},
\end{split}
\end{equation}
where
\begin{equation}
\begin{split}
&\zeta_{k}(\check{\bf P})\triangleq \check{w}_{k}\left(\left|\check{g}_{k}\right|^2\check{T}_{k}-2\mathfrak{R}\left\{\check{g}_{k}\check{\bf h}_k^H\check{\bf p}_k\right\}+1\right)-\log_2(\check{w}_{k}),\\
&\check{T}_{k}\triangleq\sum_{j\in \mathcal{K}}\left|\check{\bf h}_k^H\check{\bf p}_j\right|^2+1.
\end{split}
\end{equation}
The optimum equalizers and weights are respectively
\begin{equation}
\begin{split}\label{sharedwg}
&\check{g}_k^{*}=\check{\bf p}_k^H\check{\bf h}_k(\check{T}_{k})^{-1},\\
&\check{w}_{k}^{*}=\frac{\check{T}_{k}}{\check{T}_{k}-\left|\check{\bf h}_k^H\check{\bf p}_k\right|^2}.
\end{split}
\end{equation}\par 
We can see that \eqref{WMMSE} is non-convex because of the quadratic equality constraint, which also makes it difficult to solve. In the following part, we propose an MM-based iterative algorithm to solve this non-convex problem. \par 
At first, to reformulate the problem into a more explicit form, we define ${\bf p}_{\text{v}}=\text{vec}(\check{\bf P})$ and 
\begin{equation}
	{\bf D}_{\text{p},k}=\left[{\bf 0}^{N_{\text{t}}\times(k-1)N_{\text{t}}}\quad{\bf I}_{N_{\text{t}}}\quad {\bf 0}^{N_{\text{t}}\times(K-k)N_{\text{t}}}\right],	k\in\mathcal{K}.\label{DPK}
\end{equation}\par 
Then, the objective function in \eqref{WMMSE} can be rewritten as 
\begin{equation}
\begin{split}
	f({\bf p}_{\text{v}})
	=&{\bf p}_{\text{v}}^H{\bf Q}{\bf p}_{\text{v}}-2\mathfrak{R}\left\{\sum_{k\in \mathcal{K}}\mu_k\check{w}_k\check{g}_k\check{\bf h}_k^H{\bf D}_{\text{p},k}{\bf p}_{\text{v}}\right\}
\end{split}
\end{equation}
where 
\begin{equation}
\begin{split}
	&{\bf Q}=\sum_{j\in \mathcal{K}}{\bf D}_{\text{p},j}^H\left(\rho\sum_{k\in \mathcal{K}}\mu_k\check{w}_k\left|\check{g}_k\right|^2\check{\bf h}_k\check{\bf h}_k^H\right){\bf D}_{\text{p},j}\\&+\sum_{k'\in \mathcal{K}}\left(\frac{\rho\mu_k'\check{w}_k'}{P_{\text{t}}}(\left|\check{g}_{k'}\right|^2+1){\bf I}-{\bf D}_{\text{p},k'}^H{\bf a}(\theta_m){\bf a}(\theta_m)^H{\bf D}_{\text{p},k'}\right).
\end{split}
\end{equation}
Afterwards, \eqref{WMMSE} is equivalent to 
\begin{equation}
\begin{split}
\min\limits_{{\bf p}_{\text{v}}} \quad &f({\bf p}_{\text{v}})\\ 
s.t.\quad &{\bf p}_{\text{v}}\in\mathcal{P}
\end{split}\label{WMMSEMM}
\end{equation}
where $\mathcal{P}=\left\{{\bf p}_{\text{v}}|\text{diag}(\sum_{k\in\mathcal{K}}{\bf D}_{\text{p},k}{\bf p}_{\text{v}}{{\bf p}_{\text{v}}}^H{\bf D}_{\text{p},k}^H)=\frac{P_{\text{t}}{\bf 1}^{N_{\text{t}}\times 1}}{N_{\text{t}}}\right\}$. \par 
According to the MM framework \cite{MMTSP}, we then construct the majorization function of $f({\bf p}_{\text{v}})$. We first recall Lemma 1 from \cite{Song2015Optimization} that is \par 
\newtheorem{lemma}{Lemma}
\begin{lemma}\label{lemma1}
	Let ${\bf L}$, ${\bf M}$ be the $n\times n$ Hermitian matrices and ${\bf M}\succeq {\bf L}$. For any point ${\bf x}_0\in\mathbb{C}^n$, there is ${\bf x}^H{\bf L}{\bf x}\leq {\bf x}^H{\bf M}{\bf x}+2\mathfrak{R}\{{\bf x}^H({\bf L-M}){\bf x}_0\}+{\bf x}_0^H({\bf M-L}){\bf x}_0$.
\end{lemma}\par 
According to Lemma \ref{lemma1}, we chose ${\bf M}=\lambda_{\text{max}}({\bf Q}){\bf I}$ where $\lambda_{\text{max}}({\bf Q})$ means the largest eigenvalue of ${\bf Q}$, and have
\begin{equation}
\begin{split}
&{\bf p}_{\text{v}}^H{\bf Q}{\bf p}_{\text{v}}\\
\le&{\bf p}_{\text{v}}^H{\bf M}{\bf p}_{\text{v}}+2\mathfrak{R}\{{\bf p}_{\text{v}}^H({\bf Q}-{\bf M}){\bf p}_{\text{v}}^{t'}\}+({\bf p}_{\text{v}}^{t'})^H({\bf M}-{\bf Q}){\bf p}_{\text{v}}^{t'}\\
=&2\mathfrak{R}\{{\bf p}_{\text{v}}^H({\bf Q}-\lambda_{\text{max}}({\bf Q}){\bf I}){\bf p}_{\text{v}}^{t'}\}+2\lambda_{\text{max}}({\bf Q})P_{\text{t}}-({\bf p}_{\text{v}}^{t'})^H{\bf Q}{\bf p}_{\text{v}}^{t'}.
\end{split}\label{inequal}
\end{equation}
Here the equality is achieved at ${\bf p}_{\text{v}}={\bf p}_{\text{v}}^{t'}$. By omitting the constant items in \eqref{inequal}, we can subsequently construct the majorization function of $f({\bf p}_{\text{v}})$ as
\begin{equation}
	g({\bf p}_{\text{v}}|{\bf p}_{\text{v}}^{t'})= 2\mathfrak{R}\left\{{\bf p}_{\text{v}}^H\left[({\bf Q}-\lambda_{\text{max}}({\bf Q}){\bf I}){\bf p}_{\text{v}}^{t'}-{\bf q}\right]\right\}
\end{equation}
where ${\bf q}=\sum_{k\in \mathcal{K}}\mu_k\check{w}_k\check{g}_k{\bf D}_{\text{p},k}^H\check{\bf h}_k$. Then, \eqref{WMMSEMM} can be solved by iterating 
\begin{equation}
{\bf p}_{\text{v}}^{t'+1}= \arg\min_{{\bf p}_{\text{v}}\in \mathcal{P}}g({\bf p}_{\text{v}}|{\bf p}_{\text{v}}^{t'}).\label{MMiter}
\end{equation}
However, \eqref{MMiter} can be further investigated to find a closed-form solution. First, we denote 
\begin{equation}
\hat{{\bf q}}={\bf q}-({\bf Q}-\lambda_{\text{max}}({\bf Q}){\bf I}){\bf p}_{\text{v}}^{t'}.\label{qrep}
\end{equation}
 Then, the optimization problem \eqref{MMiter} is equivalent to 
\begin{equation}
\begin{split}
	\max_{{\bf p}_{\text{v}}}\quad &2\mathfrak{R}\left\{{\bf p}_{\text{v}}^H\hat{{\bf q}}\right\}\\
s.t.\quad & \text{diag}(\sum_{k\in\mathcal{K}}{\bf D}_{\text{p},k}{\bf p}_{\text{v}}{{\bf p}_{\text{v}}}^H{\bf D}_{\text{p},k}^H)=\frac{P_{\text{t}}{\bf 1}^{N_{\text{t}}\times 1}}{N_{\text{t}}}.
\end{split}\label{MMafter}
\end{equation}
In order to show the essence more clearly, we further denote 
\begin{align}
&\tilde{\bf q}_j=[\hat{q}_j,\hat{q}_{N_{\text{t}}+j},\dots, \hat{q}_{(K-1)N_{\text{t}}+j}]^T, j=1, 2,\dots, N_{\text{t}},\\
&\tilde{\bf p}_j=[[p_v]_j,[p_v]_{N_{\text{t}}+j},\dots, [p_v]_{(K-1)N_{\text{t}}+j}]^T, j=1, 2,\dots, N_{\text{t}},\label{trans}
\end{align}
where $\hat{q}_i$ and $[p_v]_j$ respectively denote the $i$th entry of $\hat{{\bf q}}$ and the $j$th entry of ${\bf p}_{\text{v}}$. We further define the real form as 
\begin{equation}
\begin{split}
\tilde{\bf q}^{\text{r}}_{j}=&[\mathfrak{R}\{\tilde{\bf q}_j\};\mathfrak{I}\{\tilde{\bf q}_j\}], \quad j=1, 2,\dots, N_{\text{t}}\\
\tilde{\bf p}^{\text{r}}_{j}=&[\mathfrak{R}\{\tilde{\bf p}_j\};\mathfrak{I}\{\tilde{\bf p}_j\}], \quad j=1, 2,\dots, N_{\text{t}}.
\end{split}
\end{equation}
Then \eqref{MMafter} can be reformulated as
\begin{equation}
\begin{split}
\max_{\tilde{\bf p}^{\text{r}}_{j},j=1, 2,\dots, N_{\text{t}}}\quad &\sum_{j=1}^{N_{\text{t}}}(\tilde{\bf p}^{\text{r}}_{j})^T\tilde{\bf q}^{\text{r}}_{j}\\
s.t.\quad & \lVert\tilde{\bf p}^{\text{r}}_{j}\lVert_2^2=\frac{P_{\text{t}}}{N_{\text{t}}}, j=1, 2,\dots, N_{\text{t}}.
\end{split}
\end{equation}
Following Cauchy-Schwartz inequality, we have 
\begin{equation}
\sum_{j=1}^{N_{\text{t}}}(\tilde{\bf p}^{\text{r}}_{j})^T\tilde{\bf q}^{\text{r}}_{j}\leq\sum_{j=1}^{N_{\text{t}}}\lVert\tilde{\bf p}^{\text{r}}_{j}\lVert_2\lVert\tilde{\bf q}^{\text{r}}_{j}\lVert_2=\sqrt{\frac{P_{\text{t}}}{N_{\text{t}}}}\sum_{j=1}^{N_{\text{t}}}\lVert\tilde{\bf q}^{\text{r}}_{j}\lVert_2
\end{equation}
where the last equality follows from the constraint $\lVert\hat{\bf p}^{\text{r}}_{j}\lVert_2^2=P_{\text{t}}/N_{\text{t}}$. For the condition of the equality, it is obvious that the optimal solution $\tilde{\bf p}^{\text{r}\star}_{j}$ and $\tilde{\bf q}^{\text{r}}_{j}$ should be colinear, i.e., 
\begin{equation}
\tilde{\bf p}^{\text{r}\star}_{j}=\frac{\sqrt{\frac{P_{\text{t}}}{N_{\text{t}}}}}{\lVert\tilde{\bf q}^{\text{r}}_{j}\lVert_2}\tilde{\bf q}^{\text{r}}_{j}, j=1, 2,\dots, N_{\text{t}}.
\end{equation}
Equivalently, we have 
\begin{equation}
\tilde{\bf p}^{\star}_{j}=\frac{\sqrt{\frac{P_{\text{t}}}{N_{\text{t}}}}}{\lVert\tilde{\bf q}_{j}\lVert_2}\tilde{\bf q}_{j}, j=1, 2,\dots, N_{\text{t}},\label{finalMMsol}
\end{equation}
and the solution ${\bf p}_{\text{v}}^{\star}$ can be recovered from $\tilde{\bf p}^{\star}_{j}$ according to \eqref{trans}. Then, with the WMMSE-AO framework, \eqref{joint origin} can be solved. We summarize this WMMSE-MM algorithm as Algorithm 2.  \par
\begin{algorithm}[h]
	\caption{WMMSE-MM algorithm}
	\begin{algorithmic}[1]
		\Require  $t\leftarrow0$, $\check{\bf P}^{[t]}$
		\State Calculate $\text{WSR}^{[t]}$ from $\check{\bf P}^{[t]}$; 
		\Repeat
		\State	Update ${\bf w}^{*}$ and ${\bf g}^{*}$ with $\check{\bf P}^{[t]}$ following \eqref{sharedwg};
		\State $t'\leftarrow0$;
		\State ${\bf p}^{[t']}_{\text{v}}=\text{vec}(\check{\bf P}^{[t]})$;
		\Repeat 
		\State Calculate $\hat{{\bf q}}$ with ${\bf p}^{[t']}_{\text{v}}$ following \eqref{qrep};
		\For {$j=1$ to $N_{\text{t}}$}
		\State $\tilde{\bf q}_j=[\hat{q}_j,\hat{q}_{N_{\text{t}}+j},\dots, \hat{q}_{(K-1)N_{\text{t}}+j}]^T$;
		\State $\tilde{\bf p}_{j}^*=\frac{\sqrt{\frac{P_{\text{t}}}{N_{\text{t}}}}}{\lVert\tilde{\bf q}_{j}\lVert_2}\tilde{\bf q}_{j}$;
		\EndFor
		\State Get ${\bf p}_{\text{v}}^{[t'+1]}$ using $\tilde{\bf p}_{j}^*$ by inverse operation of \eqref{trans};
		\State $t'++$;
		\Until $\lVert{\bf p}_{\text{v}}^{[t']}-{\bf p}_{\text{v}}^{[t'-1]}\lVert_2\leq\epsilon_2$
		\State  $\check{\bf P}^{[t+1]}=\text{mat}({\bf p}_{\text{v}}^{[t']})$;
		\State	Update $\text{WSR}^{[t+1]}$ using $\check{\bf P}^{[t+1]}$;
		\State	$t++$;
		\Until{$\vert\text{WSR}^{[t]}-\text{WSR}^{[t-1]}\vert\leq\epsilon_1$}
	\end{algorithmic}
\end{algorithm}
\section{Numerical Results}
In this section, we provide numerical results to validate the performance of both separated and shared transmission for joint multi-antenna RadCom system, and further reveal the advantages of shared transmission. \par 
We set that the platform adopts a ULA where $N_{\text{t}}=16$ with half-wavelength spacing, and serves $K=4$ downlink users. We assume the total transmit power budget is $P_{\text{t}}=20\text{dBm}$ and the noise power at each user is $0\text{dBm}$. Target location is set to be $\theta_m=$0\textdegree. The channel vectors of users are generated  obeying the i.i.d. complex Gaussian distribution. In the separated deployment, radar and communication subsystems fairly share the available resources, i.e., $N_{\text{tc}}=N_{\text{tr}}=\frac{1}{2}N_{\text{t}}$ and $P_{\text{c}}=P_{\text{r}}=\frac{1}{2}P_{\text{t}}$. Although radar power budget is normally higher than communication, an even power allocation here is more representative for comparison. Moreover, to ensure fair comparison between the separated and shared transmission, we set up the same channel environment in the simulations, i.e., $\check{\bf h}_k=[{\bf f}_k;{\bf h}_k]$. \par 
\subsection{Transmit beampattern comparison}
In order to evaluate the performance of transmit beamforming at the target, we first demonstrate the beampattern obtained by the separated and shared transmission.\par 
\begin{figure}[htpb]
	\centering
	\subfigure[WSR=2.4bps/Hz] {\includegraphics[width=\linewidth]{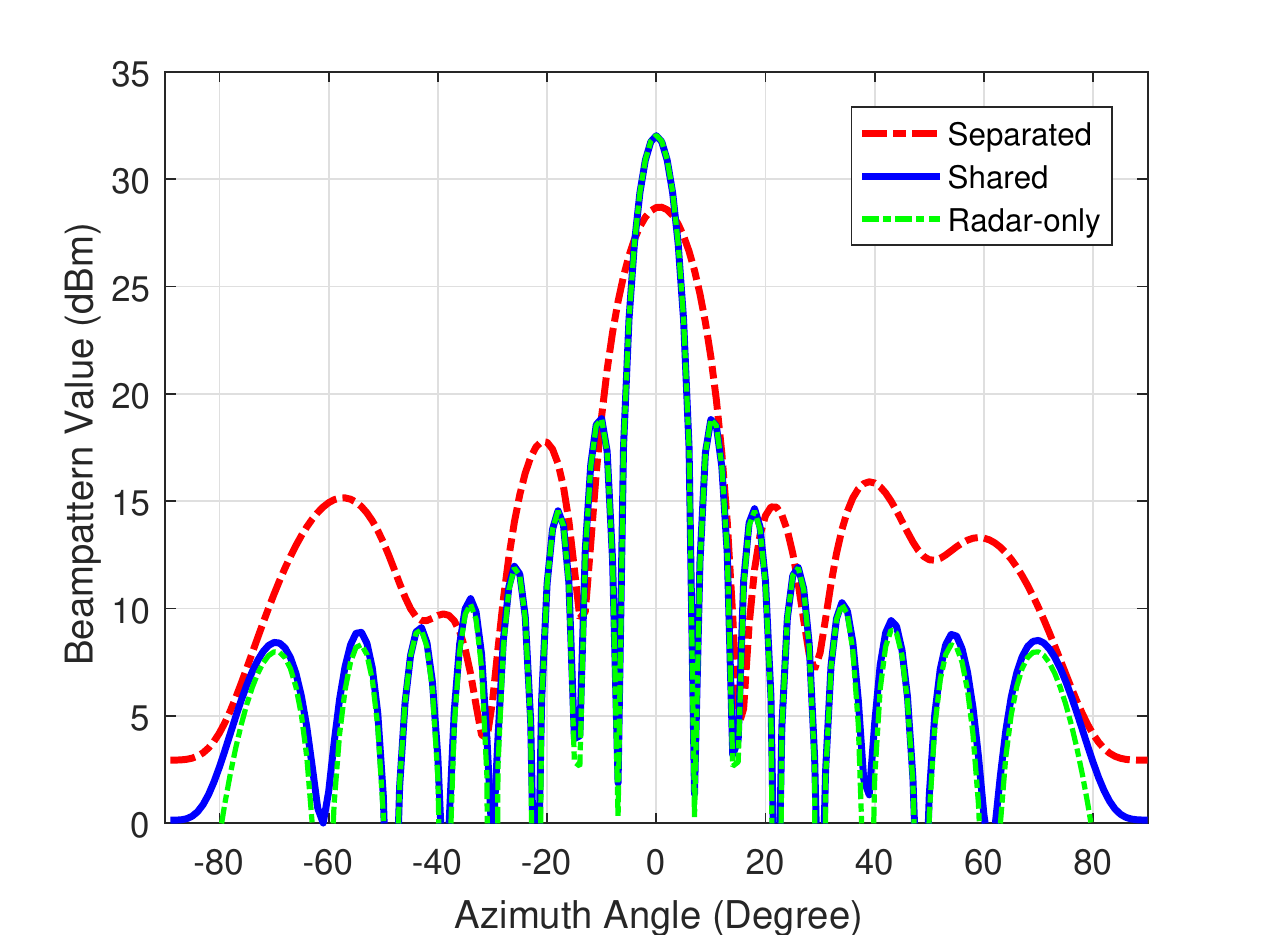}\label{low1}}
	\subfigure[WSR=5.3bps/Hz] {\includegraphics[width=\linewidth]{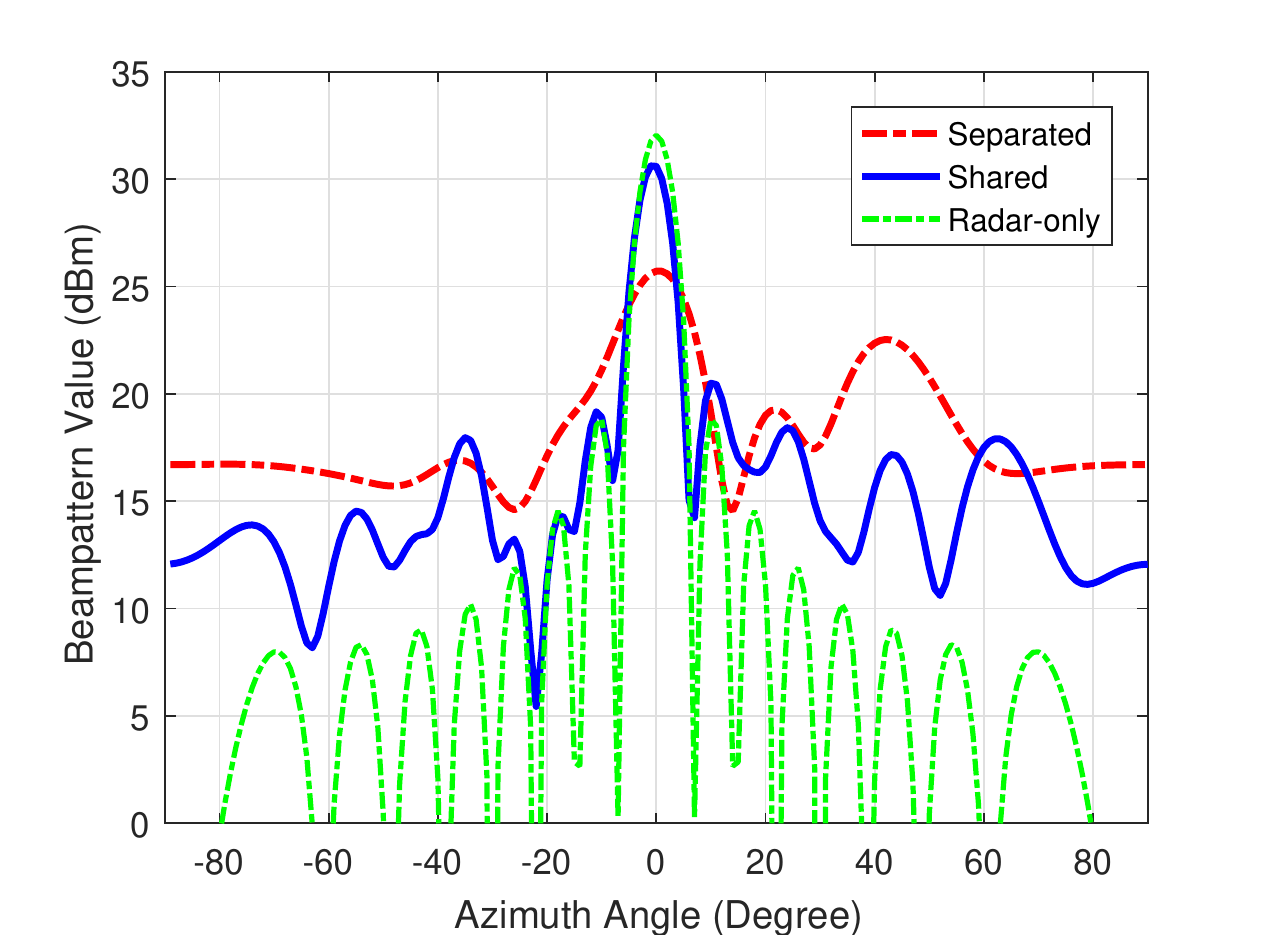}\label{high1}}
	\caption{Transmit beampattern comparison for shared and separated transmissions with different WSRs.}\label{BeamWSR}
\end{figure}
Fig. \ref{BeamWSR} compares the transmit beampattern of both transmissions obtained respectively with low and high WSRs. We can see in Fig. \ref{low1} that when WSR=2.4bps/Hz, shared transmission can nearly achieve the same beampattern as the MIMO radar equipped with the same number of antennas, showing a 3dB probing power gain at target's location over separated transmission. Fig. \ref{high1} displays that when WSR increases by 2.9bps/Hz, shared transmission experiences a 1.45dB loss of probing power at target. However, it is clear that shared transmission still keeps a 4.88dB gain over separated transmission, which is even larger than the gain in Fig. \ref{low1}. Fig. \ref{BeamWSR} also reveals that there is a tradeoff between maximizing probing power at target and maximizing WSR. \par
\begin{figure}[htpb]
	\centering
	\subfigure[WSR=1.5bps/Hz]
	{\includegraphics[width=\linewidth]{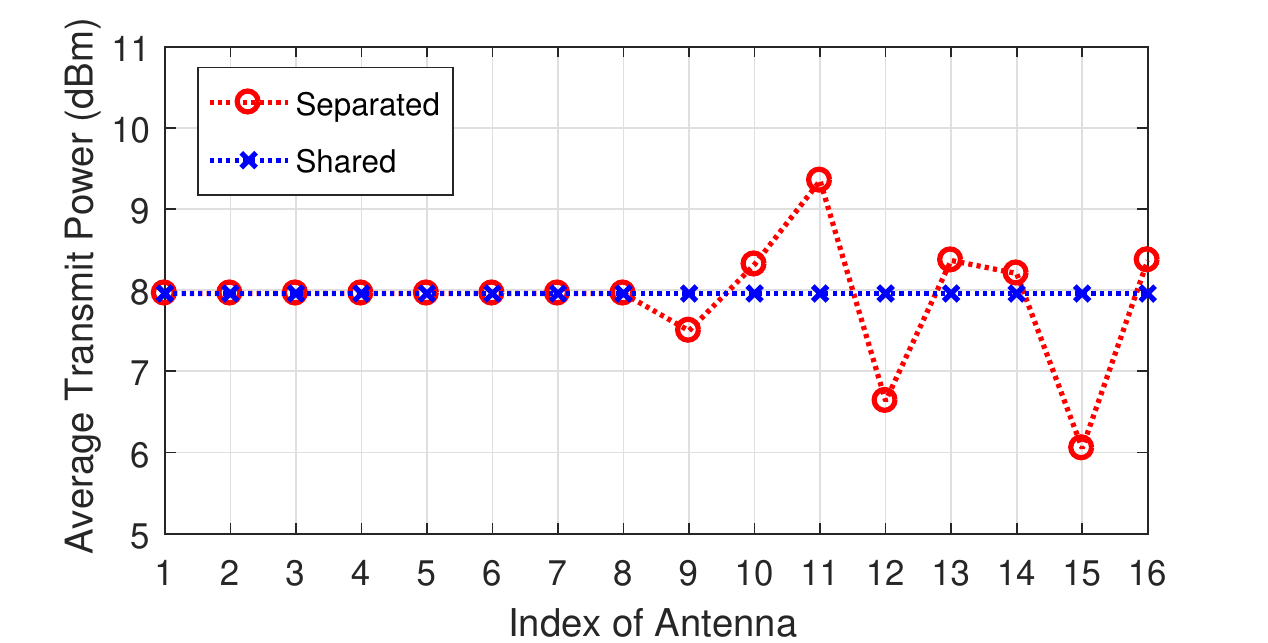}\label{perantenna1}}
	\subfigure[WSR=6.1bps/Hz]
	{\includegraphics[width=\linewidth]{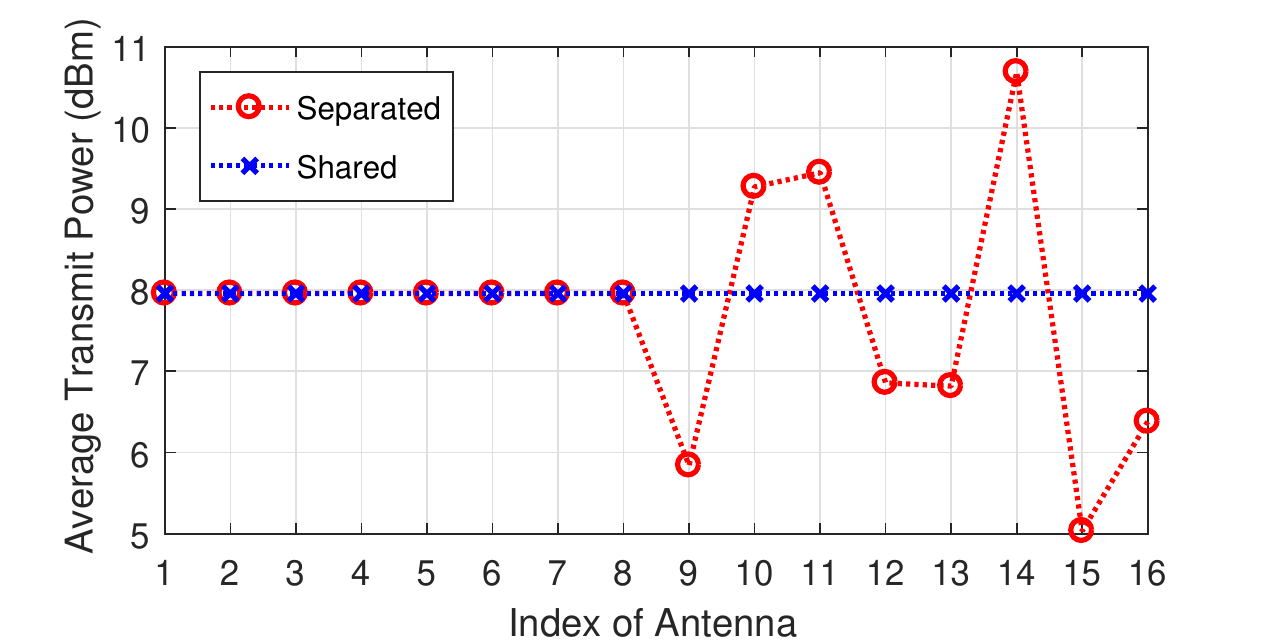}\label{perantenna2}}
	\caption{Average transmit power of each antenna for shared and separated transmission. The first eight antennas in separated transmission are intended for the radar function.}\label{perantenna}
\end{figure}

 
Fig. \ref{perantenna} shows the average transmit power at each antenna in both separated and shared transmissions corresponding to the two scenarios in Fig. \ref{BeamWSR}. Recall that the first eight antennas in the separated deployment are intended for the radar function, it is clear that the per-antenna power constraints in both shared and separated transmissions are met successfully, which proves that our algorithms handle the power constraints effectively for both transmissions.

\subsection{Tradeoff comparison}
\begin{figure}
		\centering
		\includegraphics[width=\linewidth]{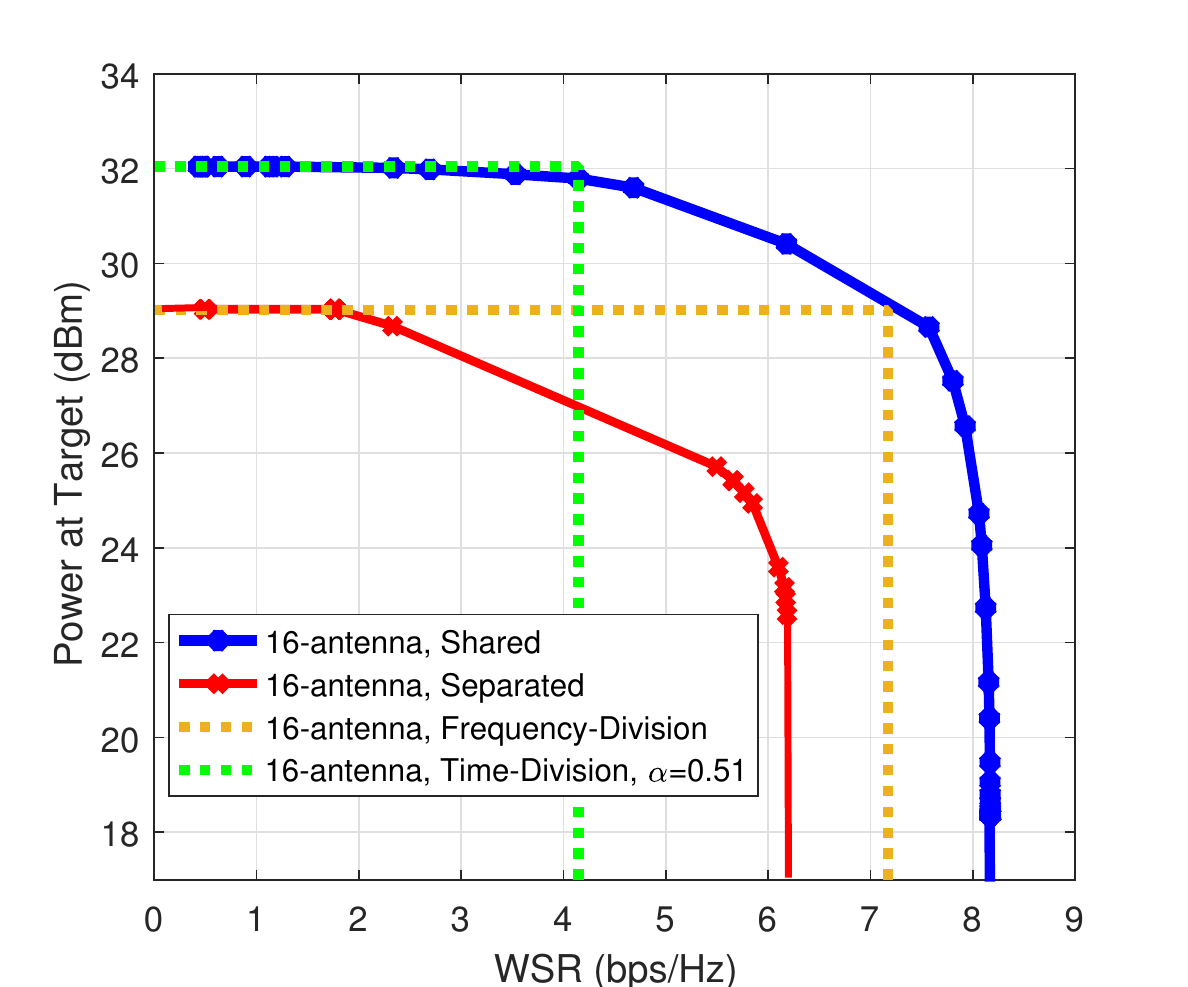}
		\caption{Tradeoff between probing power at target and WSR}\label{tradeoff}
\end{figure}

By varying the regularization parameter $\rho$ in \eqref{seperated problem} and \eqref{joint origin}, we obtain the tradeoff between WSR and probing power at target for both the shared and the separated transmissions in Fig. \ref{tradeoff} via Monte Carlo experiments. \par 
To give a well-rounded comparison, we also provide in Fig. \ref{tradeoff} two simple implementations that also achieve the dual function by orthogonalizing the resources in time (i.e. time-division) or frequency (i.e. frequency-division). \par Specifically, $N_{\text{t}}$-antenna frequency-division implementation means that a $N_{\text{t}}$-antenna system simultaneously transmits precoded communication streams and radar probing signals respectively with $P_{\text{t}}/2$ power budget but within different frequency bands. There is thus no interference between the radar and communication functions because of the frequency orthogonality. To be fair, we assume the communication precoders are optimized via SDMA based on multi-user linear precoding (MU-LP) in \cite{mao2018rate}, which maximizes WSR as well.\par 
$N_{\text{t}}$-antenna time-division implementation means a system that spends a fraction $\alpha$ of time working as a $N_{\text{t}}$-antenna BS with MU-LP and $1-\alpha$ of time working as a $N_{\text{t}}$-antenna MIMO radar with a power budget of $P_{\text{t}}$ on the same frequency band. Since radar and communication functions are realized orthogonally, WSR and probing power of both frequency-division and time-division implementations can be independently obtained by using classic methods in \cite{mao2018rate} and \cite{stoica2007probing} respectively.
\par 
In Fig. \ref{tradeoff}, we can see that separated transmission experiences a considerable performance loss as the cost of realizing spectrum sharing. To be specific, separated transmission reaches the same achievable probing power as frequency-division implementation, but sees an approximately 1bps/Hz WSR loss because it only uses half the number of antennas to transmit communication streams compared with frequency-division. Also, separated transmission shows a dual-function tradeoff because of the interference imposed by radar signals on communication users, which is a cost of sharing the same band compared with frequency division. Therefore, although separated transmission meets the RadCom requirement, it seems not to be a wise choice as the resources could be used more efficiently to improve the overall performance.\par  
In contrast, it is obvious that the shared transmission shows advantages compared with all dual-function implementations. First, it outperforms separated transmission with a maximum WSR gain of about 2bps/Hz, which results from that the former adopts twice the number of antennas and twice the power budget to transmit communication streams. We can also see that the shared transmission achieves at least 3dB gain of probing power at target given the same WSR. Second, it is clear that shared transmission surpasses frequency-division implementation with a maximum 3dB probing power gain or around 1bps/Hz WSR gain, with an additional advantage of realizing spectrum sharing. Third, as for time division, it needs pointing out that $\alpha$ varies depending on practical scenarios where the radar tracking and BS communication task are arranged based on specific demands. Therefore, for the convenience of comparison, we only provide a key baseline with $\alpha=0.51$. For larger $\alpha$, time division outperforms shared transmission, but shared transmission still has the advantage of being able to fulfill the dual function simultaneously. \par
  \par

\section{Conclusion}
To conclude, we propose two transmission design techniques that maximize WSR and probing power at target for both separated and shared RadCom deployments. We propose WMMSE-SDP and WMMSE-MM algorithms to solve the non-convex and intractable transmission design problems respectively. Numerical results show that our proposed algorithms are effective, and that shared deployment outperforms separated deployment given the same antenna number and power budget. Compared with practically simpler dual-function implementations based on time/frequency division, both separated and shared transmissions have an advantage of being able to operate the dual function simultaneously within the same frequency band. Separated transmission is less efficient in exploiting the resource and experiences a considerable performance loss compared with frequency division. In contrast, shared transmission outperforms frequency division with a maximum 3dB probing power gain or 1bps/Hz WSR gain. However, in some conditions, shared transmission is surpassed by time-division implementation but still exceeds in the capability of operating the dual function simultaneously.   
\bibliographystyle{IEEEtran}
\bibliography{IEEEabrv,xcc}

\begin{IEEEbiography}[{\includegraphics[width=1in,height=1.25in,clip,keepaspectratio]{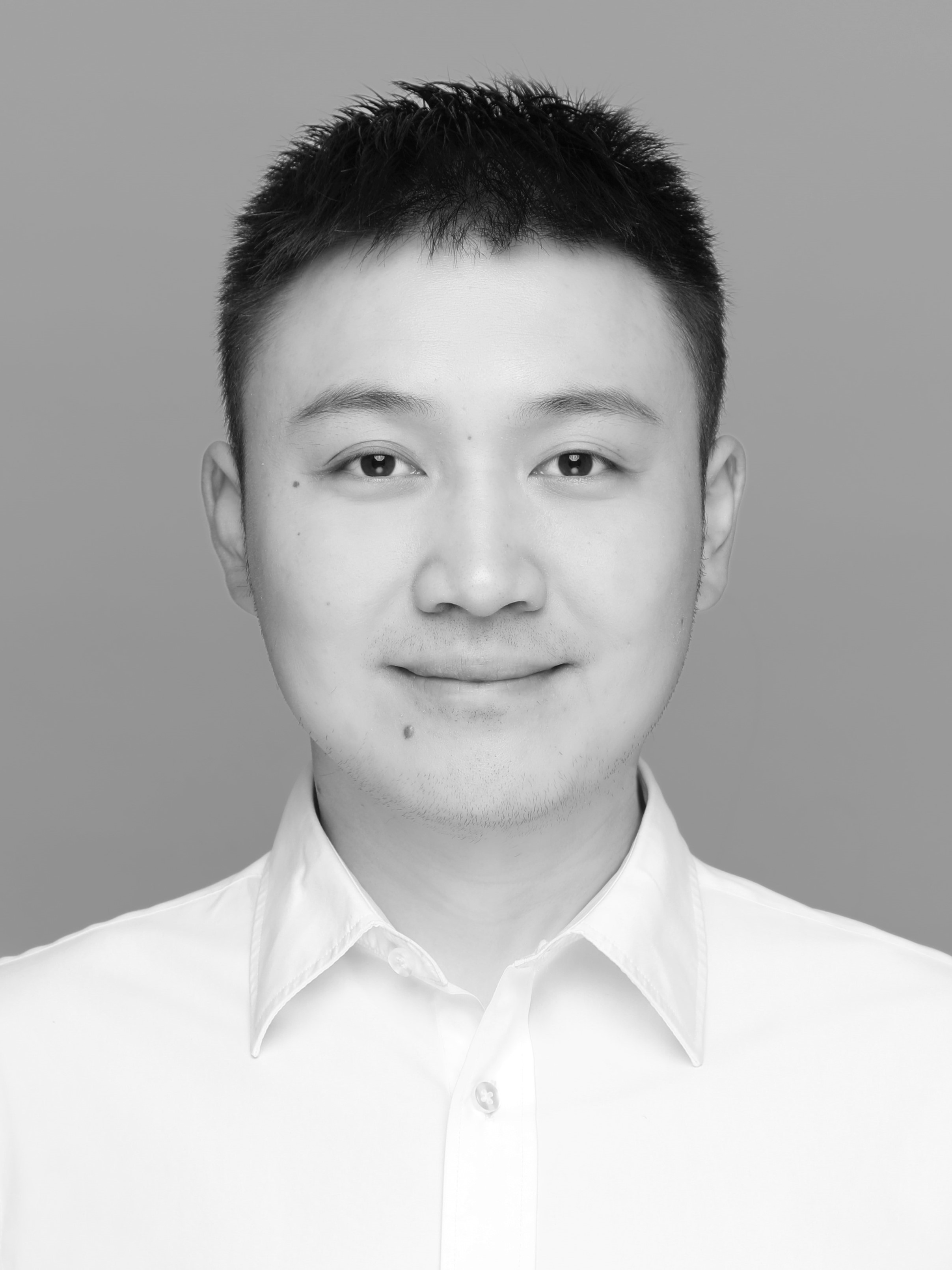}}]{Chengcheng Xu} received the master's degree in information and communication engineering from National University of Defense Technology, China, in 2017, where he is currently pursuing the Ph.D. degree with the College of Electronic Engineering. Since 2019, he has been a visiting student with the Communications and Signal Processing Group, Department of Electrical and Electronic Engineering, Imperial College London. His research interests include spectrum sensing, radar and communication spectrum sharing, and waveform design.
\end{IEEEbiography}
\begin{IEEEbiography}[{\includegraphics[width=1in,height=1.25in,clip,keepaspectratio]{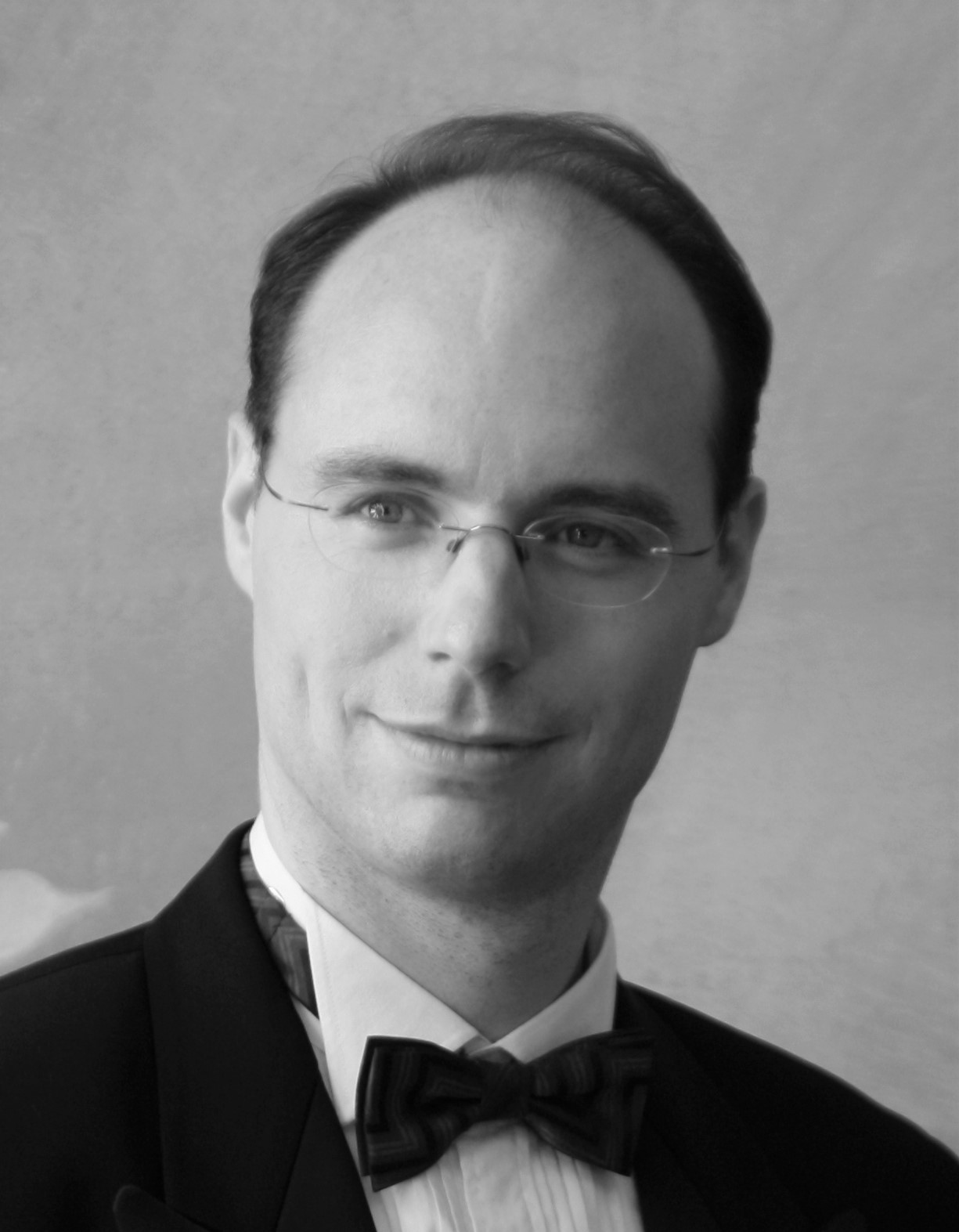}}]{Bruno Clerckx} (SM'17) is a (Full) Professor, the Head of the Wireless Communications and Signal Processing Lab, and the Deputy Head of the Communications and Signal Processing Group, within the Electrical and Electronic Engineering Department, Imperial College London, London, U.K. He received the M.S. and Ph.D. degrees in applied science from the Universit\'{e} Catholique de Louvain, Louvain-la-Neuve, Belgium, in 2000 and 2005, respectively. From 2006 to 2011, he was with Samsung Electronics, Suwon, South Korea, where he actively contributed to 4G (3GPP LTE/LTE-A and IEEE 802.16m) and acted as the Rapporteur for the 3GPP Coordinated Multi-Point (CoMP) Study Item. Since 2011, he has been with Imperial College London, first as a Lecturer from 2011 to 2015, Senior Lecturer from 2015 to 2017, Reader from 2017 to 2020, and now as a Full Professor. From 2014 to 2016, he also was an Associate Professor with Korea University, Seoul, South Korea. He also held various long or short-term visiting research appointments at Stanford University, EURECOM, National University of Singapore, The University of Hong Kong, Princeton University, The University of Edinburgh, The University of New South Wales, and Tsinghua University. \par 
	He has authored two books, 190 peer-reviewed international research papers, and 150 standards contributions, and is the inventor of 80 issued or pending patents among which 15 have been adopted in the specifications of 4G standards and are used by billions of devices worldwide. His research area is communication theory and signal processing for wireless networks. He has been a TPC member, a symposium chair, or a TPC chair of many symposia on communication theory, signal processing for communication and wireless communication for several leading international IEEE conferences. He was an Elected Member of the IEEE Signal Processing Society SPCOM Technical Committee. He served as an Editor for the IEEE TRANSACTIONS ON COMMUNICATIONS, the IEEE TRANSACTIONS ON WIRELESS COMMUNICATIONS, and the IEEE TRANSACTIONS ON SIGNAL PROCESSING. He has also been a (lead) guest editor for special issues of the EURASIP Journal on Wireless Communications and Networking, IEEE ACCESS, the IEEE JOURNAL ON SELECTED AREAS IN COMMUNICATIONS and the IEEE JOURNAL OF SELECTED TOPICS IN SIGNAL PROCESSING. He was an Editor for the 3GPP LTE-Advanced Standard Technical Report on CoMP.
\end{IEEEbiography}

\begin{IEEEbiography}[{\includegraphics[width=1in,height=1.25in,clip,keepaspectratio]{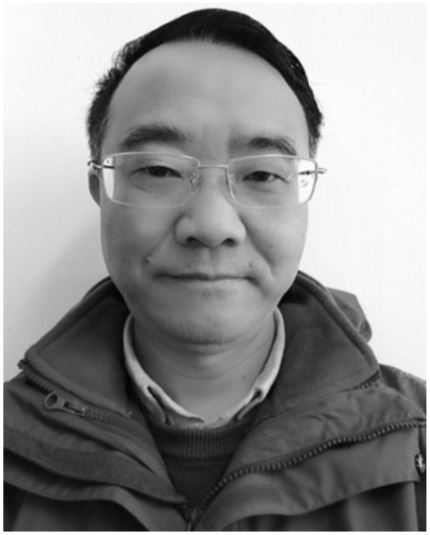}}]{Jianyun Zhang} received the B.S. degree in radar signal processing from the Electronic Engineering Institute, Hefei, China, in 1984, and the M.S. and Ph.D. degrees in digital signal processing from Xidian University, Xi'an, China, in 1989 and 1994, respectively. From 1995 to 2001, he was an Associate Professor with the Electronic Engineering Institute, where he became a Professor from 2001 to 2016. Since 2016, he has been a Professor with the College of Electronic Engineering, National University of Defense Technology. His research interests include high-speed digital signal processing, estimation theory, array signal processing, radar signal processing, and radar system theory.
\end{IEEEbiography}

\EOD

\end{document}